\documentstyle[12pt,epsf,epsfig]{article}
\def\be{\begin{equation}}
\def\ee{\end{equation}}
\def\ba{\begin{eqnarray}}
\def\ea{\end{eqnarray}}
\def\la{~\mbox{\raisebox{-.6ex}{$\stackrel{<}{\sim}$}}~}
\def\ga{~\mbox{\raisebox{-.6ex}{$\stackrel{>}{\sim}$}}~}
\def\bq{\begin{quote}}
\def\eq{\end{quote}}

 at 10truept

\newcommand{\beq}{\begin{equation}}
\newcommand{\eeq}{\end{equation}}
\newcommand{\beqa}{\begin{eqnarray}}
\newcommand{\eeqa}{\end{eqnarray}}

\def\la{~\mbox{\raisebox{-.6ex}{$\stackrel{<}{\sim}$}}~}
\def\ga{~\mbox{\raisebox{-.6ex}{$\stackrel{>}{\sim}$}}~}

\def\ltap{\ \raise.3ex\hbox{$<$\kern-.75em\lower1ex\hbox{$\sim$}}\ }
\def\gtap{\ \raise.3ex\hbox{$>$\kern-.75em\lower1ex\hbox{$\sim$}}\ }
\def\gl{\ \raise.5ex\hbox{$>$}\kern-.8em\lower.5ex\hbox{$<$}\ }
\def\roughly#1{\raise.3ex\hbox{$#1$\kern-.75em\lower1ex\hbox{$\sim$}}}

\begin{document}

\thispagestyle{empty}
\begin{flushright}
astro-ph/0511543\\ November 2005
\end{flushright}
\vspace*{1cm}
\begin{center}
{\Large \bf Of pNGB Qui${\cal N}$tessence}\\
\vspace*{1.5cm} {\large Nemanja Kaloper$^{\dagger,}$\footnote{\tt
kaloper@physics.ucdavis.edu} and
Lorenzo Sorbo$^{\ddagger,}$\footnote{\tt sorbo@physics.umass.edu}}\\
\vspace{.5cm} $^{\dagger}${\em Department of Physics, University
of California, Davis, CA 95616}\\
\vspace{.15cm} {\em $^{\ddagger}$ Department of Physics,
University of Massachusetts, Amherst, MA 01003}\\
\vspace{.15cm} \vspace{1.5cm} ABSTRACT
\end{center}
We review the pNGB quintessence models, and point out that the
reason why the large decay constants $f_a \ga {\cal O}(1)M_{Pl}$
are really needed is to tame a tachyonic instability present for a
wide range of initial $vev$s. Starting very close to potential
maxima does not help because quantum fluctuations during early
inflation at a scale $H_I$ perturb the quintessence $vev$,
displacing it from the maxima. This issue is quite interesting for
pNGB dark energy in light of the recently discussed difficulties
with embedding models with $f_a > M_{Pl}$ in fundamental theory. A
possible way around is provided by models with several ultralight
pNGBs, which can drive a short burst of very late inflation {\it
together} even if all of their decay constants obey $f_a <
M_{Pl}$. Starting with their $vev \sim f_a$, the pNGBs will hold
each other up on the potential for a longer time period. Their
effective dynamics is captured by a collective mode, containing
admixtures of all of the rolling pNGBs, which behaves as an
ultralight field with $f_{eff} > M_{Pl}$. We point out that there
may be potentially observable large scale disturbances in the sea
of dark energy in such models.

\vskip2.5cm

\begin{flushleft}
\end{flushleft}

\vfill \setcounter{page}{0} \setcounter{footnote}{0}
\newpage

\begin{flushright}
{\it ``The time has come," the Walrus said,}\\
{\it ``To talk of many things: ~~~~~~~~~~~~~~~~~~}\\
{\it Of shoes--and ships--and sealing-wax-- ~~}\\
{\it Of cabbages--and kings-- ~~~~~~~~\,~~~~~~~~~~}\\
{\it And why the sea is boiling hot-- ~~~~~~~~~}\\
{\it And whether pigs have wings." ~~~~~~~\,~~}\\

\hfil{\it (Lewis Carroll)}
\end{flushright}

There is mounting evidence that our universe may be accelerating
\cite{sne,Riess}. If gravity obeys Einstein's General Relativity
(GR), this means that the universe is currently dominated by a
dark energy component, contributing about $\sim 70\%$ to the
critical energy density. To fuel cosmic acceleration dark energy
ought to have negative pressure, satisfying roughly $w = p/\rho
\la -2/3$ to fit the observations \cite{eqst,cmblss}, if it is the
only agent responsible for supernovae dimming\footnote{See
\cite{mortost,husong} for a recent analysis of the observational
status of some alternatives.}. Because dark energy would control
both the present state of the universe and the course of its
future evolution, it is extremely interesting and important to try
to elucidate the microscopic nature of dark energy.

The first step in this quest is to devise benchmark models which
are useful to compare with observations. Dark energy is usually
modelled as a cosmological constant or as a time-dependent
quintessence field \cite{linq,q,ams,coincidence}. Each must be
finely tuned to fit the observations \cite{weinberg}. Even if we
leave aside the mystery of tuning the vacuum energy to nearly
zero, and assume that the current acceleration is due to a
quintessence field, it is hard to build realistic models.
Quintessence requires more fine tunings, over and above the scale
of dark energy density: a tiny mass $m_Q$ smaller than the current
Hubble parameter $H_0 \sim 10^{-33} {\rm eV}$ as well as
sub-gravitational couplings to normal matter, to satisfy the Solar
system gravity bounds \cite{quintconst}. This leaves but a few
models which appear natural from the point of view of 4D effective
field theory, in the sense that their masses and their couplings
to the visible matter remain small in spite of the corrections
arising  in the loop expansion of quantum field theory. In these
models quintessence is a pseudo-Nambu-Goldstone boson (pNGB)
\cite{fhsw,axq,choi,lawrence}, which in perturbation theory has
only naturally weak, derivative couplings to visible matter,
protected by a shift symmetry $\phi \rightarrow \phi + {\cal C}$.
Its potential, $V = \mu^4 \, [1-\cos(\phi/f_a)] + \ldots$, and
specifically the mass term $m_Q \simeq \mu^2/f_a$ are generated by
non-perturbative effects breaking the shift symmetry to a discrete
subgroup, and hence are radiatively stable
\cite{fhsw,axq,choi,lawrence}.

It is often quoted that to serve as quintessence, the pNGB models
need a large decay constant $f_a > M_{Pl}$ \cite{fhsw,axq}. The
idea is that the dynamics of quintessence should be akin to
chaotic inflation \cite{chaotic,abook,natural,ahchr} where a
dynamical field impersonates dark energy if the potential admits a
slow roll regime. However, for models where $f_a \ga{\cal O}(1)
M_{Pl}$ there will generically be higher order corrections to the
pNGB potential, coming as higher harmonics in the expansion over
the instanton contributions, which typically spoil the flatness of
the potential \cite{ahchr,badifogo}. Although one may invent
models where such large values of $f_a$ appear and the potential
remains flat \cite{ahchr,knp,prr}, it has been questioned whether
they can come from fundamental theory, e.g string theory
\cite{badifogo}. Even though higher order corrections might still
allow an e-fold or few of acceleration to unravel, which could be
enough to account for observations, this poses some challenges for
model building.

What of pNGBs with smaller $f_a$? One might be tempted to try to
squeeze some acceleration from them by arranging for the pNGB to
dwell on, or near, a maximum of its non-perturbatively generated
potential. Of course this entails very special initial conditions
for the pNGB $vev$, namely its initial location on the potential
peak. Also, one may try to use additional, explicit PQ symmetry
breaking terms to shift the vacuum energy, changing the Lagrangian
as in \cite{jain}, which starts with the QCD axion and chooses
$f_a \sim 10^{12} {\rm GeV}$ to argue that the axion may be dark
matter. However, in this case dark energy is really an explicit
cosmological constant term, as opposed to a dynamical
quintessence, and we have nothing new to say about it. Note also
that a pNGB with a mass $m_a \sim 10^{-16} {\rm eV}$ might weakly
mix with photons, inducing a conversion of light from distant
supernovae in the extragalactic magnetic fields, and thus also
affect the observed supernovae dimming \cite{ckt}.

In this article we will review the dynamical constraints on the
decay width $f_a$ needed for pNGB to behave as quintessence. We
will demonstrate that the simple estimates of $f_a$ based on slow
roll approximations are incomplete and somewhat misleading. The
point is that the slow roll can set in only when the field resides
in the regions of positively curved potential, $\partial^2_\phi V
> 0$. This occurs only in the region between two potential
inflections around the minimum, which is broad enough to model the
current epoch of cosmic acceleration if $f_a \ga {\cal
O}(1)M_{Pl}$. However, for models where $f_a < {\cal O}(1)M_{Pl}$,
there is no region of the convex part of the potential where the
slow roll condition $\partial_\phi V\ll V/M_{Pl}$ holds. Even if
the field starts at the inflection, the restoring force $\sim
\partial_\phi V$ will be too large and pull it down
very quickly, in time $\Delta t_{\rm inflation} \la
\frac{f_a}{M_{Pl}} \, H_0^{-1}$. Thus to mimic dark energy for
longer such a pNGB had better start near a maximum. But in this
region the pNGB is tachyonic. It grows exponentially, because its
$vev$ is sliding down the potential ridge. This tachyonic
instability is slow if $f_a \ga {\cal O}(1)M_{Pl}$, extending the
longevity of the dark energy epoch.

When $f_a < {\cal O}(1)M_{Pl}$ the instability develops much faster.
Generically the pNGB will not remain perched exactly on the top of
the potential everywhere inside the horizon even if it had started
there, because it will get displaced by quantum fluctuations
during early cosmological inflation. The amplitude of these
fluctuations sets the scale of the minimum initial displacement from the
quintessence maximum, which will be of the order of the early
inflation Hubble scale, $H_I \la 10^{13} {\rm GeV}$. For a given
$f_a < {\cal O}(1)M_{Pl}$ the field will roll down the slope fast.
Suppressing this for long enough is what places a strong bound on
$f_a$ of quintessence pNGB in this case. Further, the
perturbations imprinted in the pNGB dark energy during early
inflation will grow due to the tachyonic instability, and may lead
to potentially observable large wavelength perturbations in the
dark sector. In addition, since the pNGB $vev$ will change in the
course of cosmic evolution when $f_a \la {\cal O}(1)M_{Pl}$, the
dark energy equation of state will vary \cite{lawrence}, possibly
even leading to an increased rate of cosmic acceleration at late
times \cite{accel,Liddle}, and yielding increased supernovae
dimming at small redshifts. This offers a novel possibility of
correlating a variation of $w(z)$ with the dark energy
perturbations.

Finally, we will also point out that pNGB dynamics could still
accommodate the present epoch of cosmic acceleration even if $f_a
< M_{Pl}$ -- if there are {\it several} pNGB fields. The mechanism
which lends to this is a scaled-down assisted inflation
\cite{assisted}, whose high energy variant may emerge from string
theory \cite{nflation}. This can ameliorate some of the concerns
about the possibility of embedding pNGB quintessence with $f_a \gg
M_{Pl}$ in fundamental theory \cite{ahchr,badifogo}. Instead of a
single field pNGB quintessence, one may be able to use several
fields with $f_a < {\cal O}(1)M_{Pl}$ to dominate the universe
today. Such models may actually lead to a number of observational
signatures that could be explored by the future probes of dark
energy.

Let us first briefly review the cosmology of pNGBs. The low energy
dynamics of a pNGB is given by the Lagrangian (with the
$(-,+,+,+)$ metric convention)
\begin{equation} {\cal L} = - \frac12 (\nabla \phi)^2 - \mu^4 \Bigl(1 -
\cos(\frac{\phi}{f_a}) \Bigr) + \ldots \, . \label{lagr}
\end{equation}
The ellipsis in Eq. (\ref{lagr}) stands for the higher order
corrections from the instantons as well as possible higher
derivative terms. For now, we will ignore them. In an FRW
geometry, the homogeneous pNGB mode obeys the field equation
\begin{equation}
\ddot \phi + 3H \dot \phi + \partial_\phi V = 0 \, ,
\label{fieldeq}
\end{equation}
where $H = \dot a/a$ is the Hubble parameter and $a$ the FRW scale
factor. If the pNGB is in the slow roll regime, such that its
potential energy dominates its kinetic energy, $V > \frac12
(\partial \phi)^2$, and the restoring force in (\ref{fieldeq}) is
overwhelmed by the cosmological friction term, $\partial^2_\phi V
< 9H^2/4$, the cosmological dynamics is dominated by the
potential, yielding cosmic acceleration.

Naively this seems to occur when pNGB is displaced away from the
minimum of the potential to $\phi \simeq f_a$ (modulo $2 \pi
f_a$), such that $H^2 \simeq {\cal O}(1) \mu^4/M^2_{Pl}$ (we use
$M_{Pl} = 1/\sqrt{8\pi G_N}$ throughout) and $\partial^2_\phi V
\simeq \mu^4/f_a^2$, suggesting that as long as $f_a \ga {\cal
O}(1)M_{Pl}$ the slow roll regime may set in \cite{fhsw}. The
scale of inflation driven by pNGB depends on the scale $\mu$
generated by non-perturbative symmetry breaking
\cite{fhsw,axq,natural,ahchr,prr}, and so to generate a
sufficiently long epoch of cosmic acceleration today one must
require $\mu \sim 10^{-3} {\rm eV}$, such that $m_Q \la 10^{-33}
{\rm eV}$, as mentioned above.

However this argument is a bit too quick. Slow roll regime may
only occur if the potential is convex \cite{chaotic,abook}. Hence
the field should be within the interval around the minimum bounded
by the two adjacent inflexion points. In this regime, the slow
roll condition $M_{Pl}\,\partial_\phi V/V\ll 1$ gives
\begin{equation}
\frac{M_{Pl}}{f_a}\,
\frac{\sin\left(\phi/f_a\right)}{1-\cos\left(\phi/f_a\right)} \ll
1\,\,.
\end{equation}
If we restrict ourselves to the region for which
$V\left(\phi\right)$ is convex, the left hand side of the above
equation attains its minimum at the inflexion points
$\phi_{inflection}=\pi\,f_a/2\pm n\,\pi\,f_a$. There, its value is
$M_{Pl}/f_a$, so that one can hope to realize slow roll only if
$f_a$ is larger than $M_{Pl}$. Otherwise, the restoring force
$\sim \partial_\phi V$ will pull the pNGB down to the minimum of
the potential too quickly. Indeed: although the curvature of the
potential in (\ref{lagr}) is $\partial_\phi^2 V =
\frac{\mu^4}{f_a^2} \cos(\phi/f_a)$, and it is dominated by
friction near the inflections, the potential is well approximated
by the linear term $V \simeq \mu^4 (1- \Delta \phi/f_a)$ there.
Soon after the field is released from an initial position near the
inflection, its velocity saturates at $\dot \phi \simeq - \mu^2
M_{Pl}/(\sqrt{3} f_a)$, so that the kinetic energy of the field is
comparable to its potential unless $f_a > M_{Pl}$. This will
happen after the $vev$ of the field moves by $\sim f_a$.
Approximating the field acceleration by $\ddot \phi \sim -
\mu^4/f_a$, we can estimate the duration of an inflating phase in
this case to be $\Delta t_{\rm inflation} \la \frac{f_a}{\mu^2}$,
or therefore
\be \Delta t_{\rm inflation} \la \frac{f_a}{M_{Pl}} \, H_0^{-1} \,
. \label{inflating} \ee
Thus this regime will not last long enough unless $f_a \ga
M_{Pl}$.

What happens if the pNGB field starts near the potential maxima?
Clearly for as long as it sits precisely at the maximum, it would
behave exactly like a cosmological constant. If such a state were
very long lived, it may be able to imitate dark energy even if
$f_a < M_{Pl}$. Getting it up there clearly exacerbates fine
tuning. However one should first develop the means for quantifying
the degree of such fine tunings. One might even be able to justify
a special value of the pNGB $vev$ dynamically: e.g. in the early
universe before inflation the pNGB was a flat direction, and so it
could have fluctuated a great deal. Its $vev$ could have assumed
any values, mostly away from the value required to mimic
cosmological constant, $\phi = (\pi + 2n \pi) f_a$ for some
integer $n$. However, the required field $vev$s may have been
attained in some regions of the meta-universe simply because of
continuity, and then frozen-in by early inflation. When the pNGB
potential is generated later on, this region acquires a small
vacuum energy in the pNGB sector. So we may just happen to live in
one such domain -- inside a pNGB domain wall -- expanded to
tremendous sizes by inflation. Similar mechanisms are also used as
a catalyst of chaotic inflation \cite{chaotic} or topological
inflation \cite{topological}.

The problems with such arguments arise when one considers
dynamics. If the initial $vev$ starts inside an interval of width
$\pi f_a $ centered at maxima at $(\pi + 2 n\pi) f_a$ and bounded
by the adjacent potential inflexions, the pNGB is a tachyon: near
its maxima the curvature of the potential is negative. The pNGB
wants to slide off of the peak of the potential, and its $vev$
depends exponentially on time. When $f_a < M_{Pl}$, this time
variation is fast: the field changes exponentially in a Hubble
time. Hence in this regime, the field is {\it never} in slow roll
in the strict sense, and the epoch of dark energy domination will
quickly cease. This is really the main reason that pNGB
quintessence models work better with large $f_a$, as we will now
elaborate.

To illustrate this, we can solve the field equation
(\ref{fieldeq}) by approximating the potential as
\be V = \cases{ 2 \mu^4 - \frac12 m^2 (\phi - \phi_{max})^2 +
\ldots \, , & {\rm for} ~\, $\phi_{max} - \pi f_a/2 < \phi <
\phi_{max} + \pi f_a/2$ ; \cr \frac12 m^2 (\phi-\phi_{min})^2 +
\ldots \, , & ~~~ \, \, $\phi_{min} - \pi f_a/2 < \phi <
\phi_{min} + \pi f_a/2$ .} \label{pots} \ee
Here $\phi_{max} = (\pi + 2 n\pi) f_a$ for some integer $n$,
whereas $\phi_{min}$ is a minimum adjacent to it, where pNGB wants
to settle in, and where we have defined
\be m^2 \equiv |\partial_\phi^2 V(\phi_{extrema})| =
\frac{\mu^4}{f^2_a} \, . \label{mass} \ee
The corrections to the potential (\ref{pots}), denoted by the
ellipses, are small away from the inflections, but become
important near them since they are smoothing the potential towards
the exact cosine form. Nevertheless the corrections do not
significantly alter the qualitative aspects of the conclusions and
so we can ignore them for the most part. Now, if the $vev$ is
initially displaced from a maximum $\phi_{max}$ to $\phi_i =
(\pi/2 + n\pi) f_a + \phi_0$, where $ 0 \ne \phi_0 < f_a$ so that
$\partial^2_\phi V <0$, and if the pNGB sector dominates the
universe, the dynamics will be well approximated by $\ddot \phi +
3 H \dot \phi - m^2 \phi = 0$, with $H = \sqrt{2/3} \mu^2/M_{Pl}
\simeq {\rm const}$, with solutions $\phi \propto
\exp(\alpha_{\pm} t)$ where $\alpha_{\pm} = [-3H \pm \sqrt{9H^2 +
4m^2}]/2$. The leading order late time behavior is dominated by
the growing, tachyonic mode $\phi \propto \exp(\alpha_+ t)$. The
power $\alpha_+$ depends on the ratio of $H/m \simeq f_a/M_{Pl}$,
being\footnote{Notice that the slow roll approximation would yield
$\phi\propto\exp\left(\alpha t\right)$, with $\alpha=m^2/3H$,
that corresponds to a good approximation of the correct
result~(\ref{alpha}) only in the case $f_a>{\cal O}(1)M_{Pl}$.}
\be \alpha_+ \simeq \cases{ m \, , & {\rm for} ~\, $f_a \la {\cal
O}(1)M_{Pl}$ ;\cr
\frac{m^2}{3H} \, , & ~~~ \, \, $f_a > {\cal
O}(1)M_{Pl}$ .} \label{alpha} \ee
In either case, if the field had started at $\phi_i$ with
vanishing velocity, then at a time $t \ga 1/\alpha_+$ its $vev$
will have moved to approximately
\be \phi(t) \simeq \phi_{max} + \frac12 \, {\phi_0} \, e^{\alpha_+
t} + \ldots \, . \label{exponsoln} \ee
Once the total displacement at a time $t$ becomes of order $f_a$,
$\Delta \phi(t) \simeq \frac12 {\phi_0} e^{\alpha_+ t} \sim f_a$,
the field will have approached the inflections of the potential,
where the corrections omitted in (\ref{pots}) become important.

The effective mass $m^2_{eff} = \partial^2_\phi V$ changes along
the approach to the inflection, eventually vanishing there.
Nevertheless the kinetic energy the field accumulated on the
approach and the restoring force, $\propto \partial_\phi V \ne 0$
there, will push the pNGB through the inflection, ensuring that it
does not stop there. By this time the field velocity reaches $\dot
\phi \simeq \alpha_+ \Delta \phi(t) \simeq \alpha_+ f_a$, and so
the kinetic energy becomes $\dot \phi^2 \simeq \alpha_+^2 f_a^2$.
Depending on the ratio $H/m \simeq f_a/M_{Pl}$ the field may stop
to behave as vacuum energy by this moment. Indeed, the equation of
state parameter $w(t)$, defined as $w(t) = \frac{\dot \phi^2/2 -
V}{\dot\phi^2/2 + V}$ during this epoch grows in time
approximately as
\be w(t) \simeq -1 + \frac12 \Bigl(\frac{\alpha_+ \Delta
\phi(t)}{\mu^2} \Bigr)^2 + \ldots \, , \label{eqst} \ee
so that by the time $\Delta \phi(t) \sim f_a$, the parameter $w$
becomes
\be w_{end} \simeq -1 + \frac12 \Bigl(\frac{\alpha_+}{m}\Bigr)^2 +
\ldots \, . \label{eqstfin} \ee
Note that if $m/\alpha_+  \simeq {\cal O}(1)$, or equivalently $f_a < {\cal
O}(1)M_{Pl}$ (\ref{alpha}), $w$ will move significantly up from
$-1$ by this time, and in this case cosmic acceleration will cease
as the pNGB arrives at the inflection.

After the inflection, the field dynamics is well described by
$\ddot \phi + 3H \dot \phi + m^2 \phi = 0$. What happens to it
depends sensitively on the magnitude $f_a$. If $f_a \ga M_{Pl}$
so that $m \la H$, from (\ref{eqstfin}) it is clear that by the
time the field passes through the inflection the universe is still
accelerating (albeit marginally for $m \sim H$). By virtue of $f_a
\ga M_{Pl}$ the field next encounters a sufficiently broad swath
of convex and almost flat potential plateau, where it will in fact
dissipate at least some of its kinetic energy acquired during the
initial fast roll stage, somewhat prolonging the phase of
acceleration. In the converse case $f_a < M_{Pl}$, Eq.
(\ref{eqstfin}) shows that the cosmic acceleration has all but
stopped. In this case the potential plateau which follows the
inflection is too narrow to slow the field down. Due to its
kinetic energy, comparable to the potential at inflection, the
field will zoom by, yielding only a small fraction of an e-fold.
In either case, eventually the field will begin to oscillate
around the minimum of (\ref{pots}), obeying a well known
asymptotic attractor solution given by $H \simeq \frac2{3t}$,
$\phi = \phi_{min} + A \cos(mt +\delta)/a^{3/2}$. By virtue of the
virial theorem, the field energy density in this epoch scales as
$\rho_{\rm pNGB} \simeq \mu^4 (\frac{a_0}{a(t)})^3$, where $a_0$
is roughly the value of the scale factor when the epoch of
acceleration ended. Depending on the scales however, it may yet
occur that after the first pass through the minimum the field
continues to climb up the potential past the inflection on the
opposite side. In that case, the field would yield another short
burst of inflation as it slowly climbs the opposite slope, before
returning to oscillations around the minimum.
The possibility of having inflation briefly start again after the
passage through the minimum has been considered earlier in
relation to the cosmological moduli problem \cite{thisla}. Such
dynamics could simulate the behavior of dark energy with $w<-1$,
as recently discussed in \cite{accel,Liddle}.

To summarize this discussion, we see that a universe controlled by
a pNGB may first undergo a stage of inflation during which the
field changes quickly, akin to the fast-roll inflation of
\cite{fastroll}, which may (for $f_a \ga M_{Pl}$) or may not (for
$f_a < {\cal O}(1) M_{Pl}$) be followed by a stage of slow roll
inflation, and then eventually end up like a universe dominated by
an oscillating massive scalar field. If
$m/\alpha_+ \simeq {\cal O}(1)$,
or therefore $f_a \la {\cal O}(1) M_{Pl}$ (\ref{alpha}),
cosmic acceleration will come to a close by the time the pNGB $vev$
gets to the inflection. In this case the duration of the
accelerating epoch is fully controlled by the coefficient
$\alpha_+ \sim m$ (see Eq. (\ref{alpha})) and the initial
displacement from the maximum,
\be \Delta t \sim \frac1{m} \ln\Bigl(\frac{f_a}{\phi_0}\Bigr) \, .
\label{duration} \ee
If a single pNGB is to
mimic dark energy today, $\Delta t$ must be of the order of the
current age of the universe, $\sim 1/H_0$. To ensure this the
initial displacement of the pNGB $vev$ away from the potential
maximum should be no more than
\be \phi_{0~max} \sim f_a e^{-m/H_0} \sim  f_a e^{-M_{Pl}/f_a} \,
, \label{initial} \ee
to ensure that the universe is accelerating now.

Herein lies the problem: the condition $\phi_0 < \phi_{0~max}$
will not hold unless $f_a \ga M_{Pl}$. By its very definition,
pNGB couplings are very weak and its potential, generated by
non-perturbative effects, is very flat\footnote{In fact at
energies above the scale of shift symmetry breaking, there is no
potential for pNGB at all.},
compared to a typical scale of early
inflation. Therefore during early inflation which shaped our
universe and generated primordial perturbations, the pNGB field is
relativistic, with $\partial^2_\phi V \ll H^2_I$. Hence like any
relativistic field during inflation it must have fluctuated a lot
around its mean value. Its dynamics was stochastic
\cite{abook,selfrepr}: imagine that in some region of the early
universe the pNGB $vev$ did attain exactly the value corresponding
to the maximum of the potential in (\ref{lagr}), $\phi = (\pi +
2n\pi) f_a$, when inflation started. Because it is dynamical and
relativistic, the field subsequently gets displaced by the quantum
fluctuations generated in de Sitter environment
\cite{abook,selfrepr,gibhawk,stayo}. It ``wiggled" a little inside
the inflating region from one Hubble time to another. The scale of
these quantum fluctuations is given by the root mean square value
of the field around its expectation value, set by the two point
function of a relativistic quantum field in de Sitter space
\cite{abook,selfrepr,stayo}:
\be < \phi^2(t)> \,= \, \Bigl(\frac{H_I}{2\pi}\Bigr)^2
\int^{H_I}_{H_I e^{-H_It}} \frac{d^3 \vec k}{k}  \, . \label{rms}
\ee
From this, the jump of the field between two subsequent Hubble
intervals is
\be \delta \phi \simeq \frac{H_I}{2\pi} \, . \label{jump} \ee

The constraints on $f_a$ when it is smaller than $ {\cal O}(1)
M_{Pl}$, derive from this. While dependent on the scale of early
inflation, they are nevertheless useful. If the field is initially
displaced by $\sim H_I/2\pi$ (\ref{jump}), due to the tachyonic
instability it will slide off the maximum in time $\Delta t$ given
by (\ref{duration}). If this is to yield enough late cosmic
acceleration, $\Delta t  \ga 1/H_0$. Thus we find
\be M_{Pl} \la f_a \ln\Bigl(\frac{2\pi f_a}{H_I} \Bigr)  \, ,
\label{timeconst} \ee
where we used (\ref{mass}) and $H_0 \sim \mu^2/M_{Pl}$. The
inequality (\ref{timeconst}) implies that the logarithm ranges
between $\sim 15$ and $\sim 36$ for the range of $H_I$ from about
$10^{13} GeV$ down to about $10^4 GeV$, which cover many generic
inflationary models including chaotic inflation \cite{chaotic} and
hybrid inflation \cite{hybrid}, its supergravity realizations
\cite{sugr} and modern hybrid inflation models on the string
landscape \cite{kklt}. Using (\ref{timeconst}) we can estimate
that $f_a$ must
be greater than about a tenth of $M_{Pl}$ if pNGB
is to mimic dark energy today:
\be f_a \ga 0.1 \times M_{Pl} \, . \ee
In fact, numerical considerations show that the bound is even
stronger by a factor of a few.

\begin{figure}[thb]
\centerline{\epsfbox{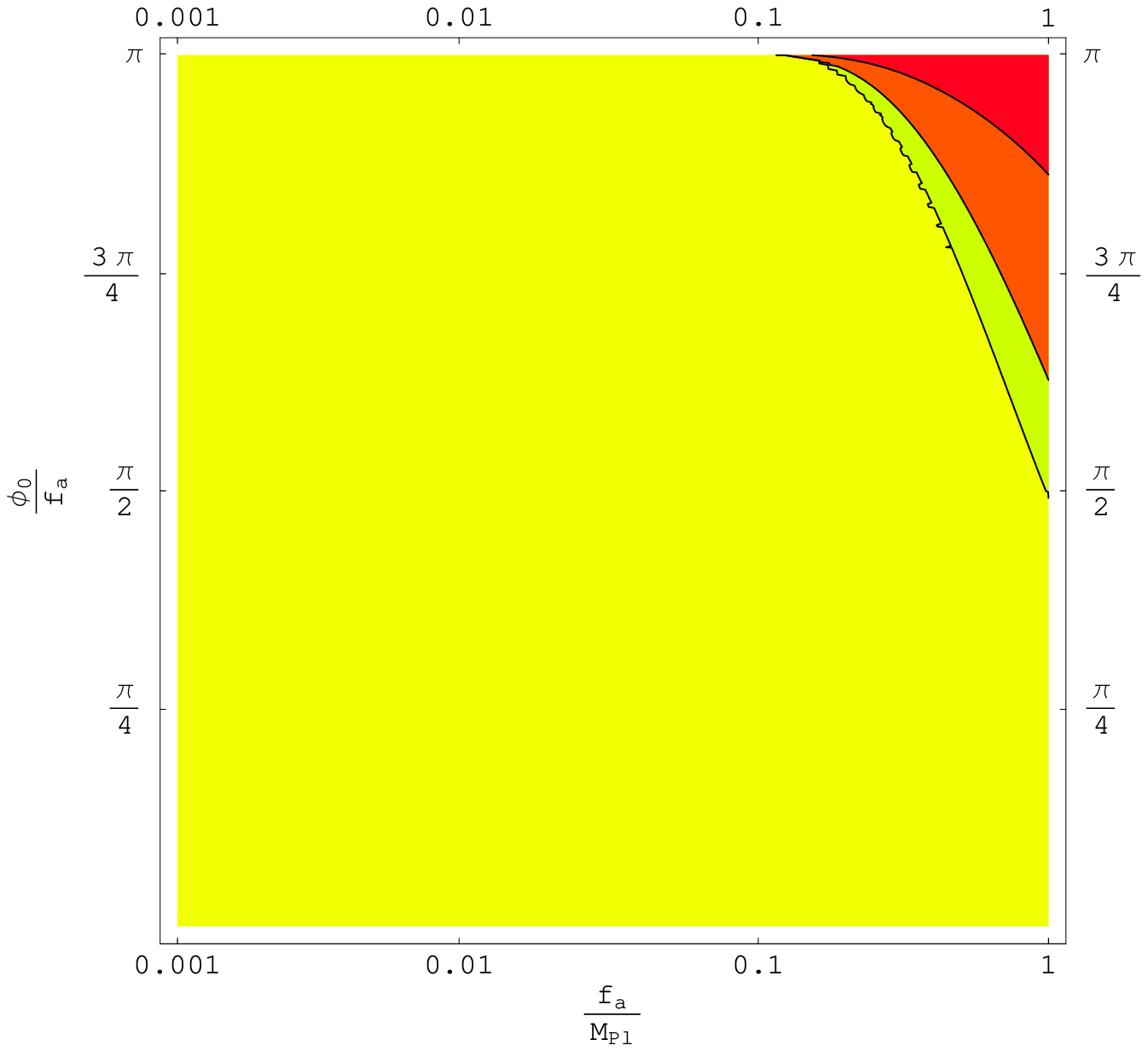}} \caption{Current value of the
equation of state parameter $w_0$ of the pNGB. Going inwards from
the top right corner of the figure, the different colors
correspond to areas with $-1<w_0<-0.99$ (red), $-0.9<w_0<-0.99$
(orange), $-0.7<w_0<-0.9$ (green) and $w>-0.7$
(yellow).\label{param}}
\end{figure}

In Figure~\ref{param} we plot the current value of the equation of
state parameter of the pNGB, as a function of its initial value
$\phi_0$ and of $f_a/M_{Pl}$. Current observations relegate the
allowed part of parameter space to the top right corner of the
figure, and -- barring fine tuned initial conditions -- a typical
bound on $f_a$ reads $f_a > 0.2\,M_{Pl}$, to fit the observations.
We should remark, here, that the boundary between the forbidden
light-colored region and the allowed darker one is not as sharp as
it appears on the plot. Indeed, for values of $\phi_0$ and $f_a$
close to the boundary, we have regions where the pNGB has
performed a half oscillation around the minimum before the present
time, and has been recently {\em climbing} its
potential\footnote{These regions are not represented in
Figure~\ref{param}, because in the plot we have computed the value
of $w$ at the moment in  which $\Omega_{\mathrm {matter}}$ crosses
the value $0.3$ for the first time. In the "forbidden" region, the
value of $w$ is larger than $-0.7$ when this first crossing
happens. But for values close to the boundary between the allowed
and the forbidden region, $\Omega_{\mathrm {matter}}$ actually
performs small oscillations around the value $0.3$, while the
equation of state parameter $w$ of the pNGB oscillates between
$-1$ and $1$.}. In such cases, the initial potential energy of the
field starting near the maximum does not dissipate as quickly, and
they swing past the minimum beyond the opposite inflection, where
they linger a little on the potential. 
The study of the viability of these regions is more complicated, as the full time dependence of the parameter $w(a)$ should be taken into account in this case, along the lines of~\cite{wa}.
These regions could
actually provide a better fit to supernova data, mimicking
$w<-1$~\cite{accel, Liddle}.

We can thus conclude that models where $f_a<0.1\,M_{Pl}$ are {\it
extremely} fine-tuned: to get a pNGB to mimic dark energy when
$f_a$ fails to satisfy (\ref{timeconst}) one needs {\it extreme}
fine tunings of the initial state of the pNGB. The field must
start in a special state where the subsequent quantum fluctuations
precisely cancel against the prearranged initial dark energy
inhomogeneities. This is extremely unlikely, and goes against the
grain of the standard inflationary lore. In inflation, such
special features of the initial state would be erased by
exponential expansion, and the subsequent structure would be
imprinted by quantum fluctuations, leading to (\ref{timeconst})
for pNGBs with $f_a < M_{Pl}$.

If we parameterize the end of inflation by picking a spacelike
surface along which the adiabatic perturbations generated during
inflation are $\delta \rho/\rho \simeq 10^{-5}$, we will find that
the distribution of the $vev$s of the pNGB field is also slightly
inhomogeneous, given by an approximately scale invariant spectrum
at scales $\ga 1/H_I$ and with the field perturbation estimated
by, roughly, Eq. (\ref{jump}). These perturbations of the pNGB
$vev$ will remain frozen until the horizon reentry at a much later
time. They will yield small initial perturbations of the dark
energy density,
\be \frac{\delta \rho_{DE}}{\rho_{DE}} \simeq \frac{\delta V}{V}
\simeq \frac{H^2_I}{8\pi^2 f_a^2} \, , \label{deperts} \ee
where we assume that the field was initially exactly on top of the
potential. In evaluating the last line of (\ref{deperts}) we have
used (\ref{mass}). If $\phi$ started off farther from the maximum,
the perturbations would be larger,
\be \frac{\delta \rho_{DE}}{\rho_{DE}} \simeq \frac{H_I}{2\pi f_a}
\, . \label{depertoff} \ee
Initially, the dark energy perturbation may be very small. For a
typical inflationary model with $H_I \la 10^{13} {\rm GeV}$, if
the pNGB decay width is large, $f_a \ga M_{Pl}$, the initial dark
energy perturbation would be ${\delta \rho_{DE}}/{\rho_{DE}} \la
10^{-14} - 10^{-7}$, subsequently slowly growing in time, as the
pNGB slides off the maximum. When $f_a < {\cal O}(1) M_{Pl}$,
however, the perturbation would both be initially larger and would
grow faster. Indeed, in regions of the postinflationary universe
where inflation displaced pNGB away from the maximum by $\sim H_I$
the late stage of acceleration driven by the pNGB would terminate
faster than in those regions where the field remained perched on
top of the potential. Such models can develop large perturbations
at late times in the dark sector, because the initially small dark
energy density perturbations will grow as pNGB slides down the
potential
\footnote{
For some choice of parameters, the fluctuations might backreact on the evolution of the zero mode of the pNGB~\cite{ch}, affecting its equation of state.}.
Even if they start as small as (\ref{deperts}) or
(\ref{depertoff}), by the time the field approaches the potential
inflection, $\Delta \phi \rightarrow f_a$, the perturbations will
increase according to
\be \frac{\delta \rho_{DE~end}}{\rho_{DE~end}} \simeq
\Bigl(\frac{\Delta \phi(\Delta t)}{f_a}\Bigr)^2 \rightarrow {\cal
O}(1) \, , \label{endperts} \ee
increasing as $f_a$ drops below $M_{Pl}$. Thus the dark energy
perturbations could in principle grow to become sufficiently large
today to compete with the inflationary perturbations at large
scales. We expect that these dark energy perturbations would yield
isocurvature perturbations at very large scales and very late
times, and the mechanism would be akin to the curvaton mechanism
of \cite{curvaton}. While here we will refrain from a detailed
computation of the spectrum and evolution of dark energy
perturbations, we note that such a study has been performed on
pre-WMAP data in~\cite{kmt}
(a previous study of this issue can be found in~\cite{vl}, where
the authors,  however, focused only on the special value
$f_a=M_{Pl}$).
The constraints found in~\cite{kmt}
are  weaker than the ones imposed by supernovae. The reason,
roughly, is that during inflation structure grows slowly, and
hence the perturbations remain frozen until the end of the
accelerating epoch. An analysis that includes the more accurate
WMAP data may help reducing the available portion of the parameter
space,\footnote{See \cite{cpkl,lowq} for an interesting
discussion of the possibility that such perturbations might play a
role in explaining the observed lack of power in low-$\ell$ CMB
multipoles.} although it should also be clear that if $f_a \ga
{\cal O}(1) M_{Pl}$ such effects may remain small\footnote{When
$f_a \ga {\cal O}(1) M_{Pl}$ the pNGB $vev$ will change more
slowly in a Hubble time $\sim 1/H_0$, and also the field need not
start from the maximum of the potential.}. The possibility of
generating and observing the imprints of early inflation in dark
energy has been recently discussed in more general terms in
\cite{stochastquint}.

These arguments show that the pNGB quintessence models need $f_a
\ga {\cal O}(1) M_{Pl}$ to be viable. How could such models arise
from fundamental theory, given the problems so far with deriving
large $f_a$ in string theory \cite{ahchr,badifogo}? In fact, there
is a simple way to provide for this using pNGBs with $f_a < {\cal
O}(1) M_{Pl}$, if there is many of them\footnote{We exploit the 
possibility of having many pNGBs with
couplings which differ from each other, in contrast to the case of
{\it two} pNGBs of \cite{knp} where an accidental symmetry
relating axion couplings was used to generate a large $f_{eff}$.}. It is based on the
mechanism of assisted inflation \cite{assisted,nflation}. Imagine
that there are ${\cal N}$ copies of a pNGB with $f_a < {\cal O}(1)
M_{Pl}$, which interact with each other via gravity. Then the
total Lagrangian of the system is the sum of ${\cal N}$ copies of
(\ref{lagr}). Assume for simplicity that all the fields are
displaced away from the minimum by the same amount. In that case,
we can introduce a collective coordinate
\be \varphi = \sqrt{\cal N} \phi \, , \label{collphi} \ee
whose effective potential is given by
\be V_{eff} = {\cal N} \mu^4
\Bigl(1-\cos(\frac{\varphi}{\sqrt{\cal N} f_a}) \Bigr) \, .
\label{effpot} \ee
Thus we see that the collective coordinate $\varphi$ behaves as a
single pNGB with a potential controlled by the symmetry breaking
scale $\mu_{eff}$ and decay width $f_{eff}$
\ba \mu_{eff} &=& {\cal N}^{1/4} \mu \, , \nonumber \\
f_{eff} &=& {\cal N}^{1/2} f_a \, . \label{scales} \ea
Therefore the total vacuum energy and the effective mass analogous
to (\ref{mass}) for the collective mode $\varphi$ are
\ba \lambda_{eff} &=& {\cal N} \mu^4 = {\cal N} \lambda \, ,
\nonumber \\
m^2_{eff} &=& \frac{{\cal N} \mu^4}{{\cal N}^2 f_a^2} =
\frac{m^2}{\cal N} \, . \label{params} \ea
We can therefore approximate the potential for $\varphi$ with an
expression just like (\ref{pots}), where we replace $\mu$, $m$ and
$f_a$ by $\mu_{eff}$, $m_{eff}$ and $f_{eff}$ given in
(\ref{scales}), (\ref{params}). The analysis of the cosmological
dynamics controlled by the collective mode $\varphi$ then is the
same as the investigation of the cosmology dominated by a zero
mode of a single field $\phi$, and we see that by virtue of Eqs.
(\ref{scales}), (\ref{params}) the duration of the epochs of dark
energy is prolonged. If the field $\varphi$ starts below the
inflection, where the potential is convex, the duration of the
slow roll regime is approximately the same one that we would find
in a simple quadratic potential. Indeed, the field $\varphi$ will
be slowly rolling until its $vev$ is at a distance of the order of
the Planck mass from the minimum of $V\left(\varphi\right)$. The
total number of e-folds will be ${N} \simeq f_{eff}^2/M_{Pl}^2
\sim {\cal N}$, which we can again estimate using the formula
relating the number of e-folds $N$ and the field $vev$ for a
polynomial potential $V \sim \lambda \varphi^n/n$, $N = (\varphi^2
- \varphi_0^2)/(2nM^2_{Pl})$ \cite{chaotic,abook,kkls}. Assuming
${\cal N} > (M_{Pl}/f_a)^2$ we can approximate the potential by
the quadratic term, as opposed to the linear potential as in the
case leading to Eq. (\ref{inflating}). This yields
\be \Delta t_{\varphi ~{\rm roll}} \sim \frac{f^2_{eff}}{M^2_{Pl}}
\, H_0^{-1} \sim {\cal N} \frac{f^2_a}{M^2_{Pl}} \, H_0^{-1}  \, .
\label{phirol} \ee
Notice that this is the regime considered, in the context of
primordial inflation, in~\cite{nflation}. Likewise, the tachyonic
instability is also slowed down. When ${\cal N} > M_{Pl}^2/f_a^2$
so that $f_{eff} > M_{Pl}$, the value of $\alpha_+$ in
Eq.~(\ref{alpha}) is now given by $\alpha_+\simeq m_{eff}^2/3H$,
and so the Eq.~(\ref{duration}) is now replaced by
\begin{equation}
\Delta t_\varphi \sim
\frac{3H_0}{m_{eff}^2}\,\ln\left(\frac{f_{eff}}{\varphi_0}\right)
\sim {\cal N} \, \frac{3H_0}{m^2} \ln\Bigl(\frac{\sqrt{\cal N}
f_{a}}{\varphi_0}\Bigr) \,\,. \label{newdur}\end{equation}
Again, it takes $f_{eff}^2/M_{Pl}^2$ e-folds to get to the
inflection. But now even if $f_a/M_{Pl} < 1$, as long as there are
many pNGBs such that $\sqrt{\cal N} f_a/M_{Pl} > 1$, a full e-fold
of acceleration or more may ensue.

Obviously it is important that the pNGBs are not directly mutually
interacting. Direct interactions are detrimental to the effect of
assisted inflation, which can be seen very simply from the
following argument. The effect of assisting works by taking a
given potential of fixed curvature and then prolonging the epoch
of dark energy supported by this potential by taking ${\cal N}$
degenerate copies \cite{assisted}. If the direct interactions are
then turned on, they will lift the degeneracies of the mass
spectrum of the fields. Some fields will be lighter, but most will
get heavier. This is akin to a generic effect of mass splitting of
a degenerate quantum system under a perturbation. Hence, in spite
of a large number of fields, some of them will become so heavy
under the interactions that they will not be able to stay on the
potential ridge, and will tumble down. However, it may be possible
to ensure that the mutual interactions are miniscule by arranging
for the right compactification of string theory, where pNGBs are
axion-like degrees of freedom present in the moduli space, along
the lines of \cite{nflation}. A version of such a mechanism with
very low mass scales may find its home in the landscape of string
solutions.

Our argument that many copies of pNGBs  prolong the epoch of dark
energy relative to a single one is clearly sensitive on the
initial conditions for the individual pNGBs in the early universe.
The assisted dynamics works best if the pNGBs start out at roughly
the same place, away from the minimum of the potential, where they
hold each other up on the potential slope. In general, however,
the effective field $\varphi$ will have some width, since the
pNGBs may start at a different place. Further, they may also have
slightly different effective masses (even if their $\mu$ and $f_a$
are all the same, the effective mass is determined by the
curvature of the potential and hence depends on the position of
the field). Because of this, the field $\varphi$ will eventually
begin to fall apart in the course of evolution, and this will
shorten the dark energy epoch and may lead to a new source of dark
energy perturbations. While it is obvious that tuning the initial
values of the separate pNGBs will hold this under control, it
would be interesting to revisit this issue in more detail.

Finally, let us comment on the idea that the pNGBs might have so
extended a spectrum of masses and couplings that some of them
contribute to dark energy today while others have already been
oscillating for a long time around the minimum of their potential,
behaving in effect as dark matter. In this scenario, dark matter
and dark energy would share a common origin, a fact that might
help with the coincidence problem. We find that such mechanisms
are hard to implement naturally in a simple setting. To illustrate
these difficulties, consider the simplest case of two pNGBs and
ask that one of them (with parameters $\mu_1$ and $f_1$) be dark
matter, while the other ($\mu_2$, $f_2$) be dark energy. The first
pNGB will start rolling along its potential when the Hubble
parameter $\sim T^2/M_{Pl}$ equals its mass $\mu_1^2/f_1$. This
happens at a temperature $T_1\sim \mu_1\sqrt{M_{Pl}/f_1}$. The
pNGB subsequently evolves as matter, so that
\begin{equation}
\rho_1\simeq \mu_1^4\,\left(\frac{T}{T_1}\right)^3\,\,,
\end{equation}
and it starts dominating the energy density of the Universe at the
temperature $T_e\sim 1$~eV of matter-raidation equality.
We thus find the relation
$\mu_1\simeq T_e\left(M_{Pl}/f_1\right)^{3/2}$. By requiring
$T_1>T_e$, we get $f_1<M_{Pl}$, that implies $\mu_1>T_e$. The
parameters of the second pNGB are determined by the requirement
the it is dark energy today: $\mu_2\simeq 10^{-4} T_e$, $f_2\simeq
M_{Pl}$. To solve the coincidence problem, one would need to
relate the two pNGBs in some natural way, requiring that their
mass scales and couplings $\mu_k, f_k$ are related by a common new
physics at some new scale $\Lambda$. Parameterizing this
relationship by a simple power law, so that
$\mu_{1,2}=\Lambda^{1-\beta}\,f_{1,2}^\beta$, we get the relation
\begin{equation}
\left(f_1/M_{Pl}\right)^{3/2+\beta}=10^4\,\,.
\end{equation}
As $f_1<M_{Pl}$, $\beta$ has to be negative, and $\mu$ should be
thus inversely proportional to some power of $f$.  Typically
models of pNGB have the opposite behavior, and so this would
require some complicated and unusual construction to reproduce the
desired spectrum of pNGBs, pursuit of which is beyond the present
work.

To recapitulate: we have revisited models where dark energy is
dynamical, controlled by a pNGB field. We have stressed that
unless $f_a \ga {\cal O}(1)M_{Pl}$, models with a single, or very
few, pNGB(s) are not good explanations of the present epoch of
dark energy domination because of the tachyonic instability
plaguing most of the field initial values. Even if the field
starts out very close to a maximum of its potential, because it is
so light quantum fluctuations during early inflation will shift it
away from the maximum by an amount proportional to the Hubble
scale of inflation $H_I$. This would exacerbate the fine tuning
problems of dark energy. Given the concerns raised about embedding
pNGB models with $f_a > M_{Pl}$ in fundamental theory
\cite{ahchr,badifogo} on the one hand, and the acute absence of
technically natural models of quintessence on the other
\cite{ahchr,quintconst}, this may present a challenge for building
dark energy models within the realm of effective field theory
consistent with fundamental theory. We have noted that a possible
solution of this problem can be developed by scaling down the
mechanism of assisted inflation \cite{assisted,nflation} to very
low scales. If there are several ultralight pNGBs, they may drive
very late epoch of cosmic acceleration together even if their
individual decay constants satisfy $f_a < M_{Pl}$. We should note
that although the assisted inflation idea does {\it not} seem to
help other quintessence models \cite{kilitsu}, it works well for
pNGBs. If quintessence is really an admixture of many modes, its
decay may leave signatures in the form of inhomogeneities of dark
energy distributions. These inhomogeneities are triggered by
quantum fluctuations during early inflation, just as the primordial
density perturbations, and ultimately baby universes in
self-reproducing inflationary models \cite{selfrepr}. Thus in some
sense such perturbations may carry information about the cosmic
seeds, frozen in the cosmic permafrost that is dark energy. It
would be interesting to carry out more precise calculations of the
amplitude of these perturbations and check their observational
consequences, since they may affect cosmological anisotropies at
the largest of scales \cite{cpkl,lowq}. Further, in such models
dark energy dynamics may also result in a variable equation of
state $w(z)$, which could account for additional dimming of
supernovae at low redshifts \cite{accel,Liddle}, warranting
coordinated searches for $w'$ and for dark energy perturbations.

Finally, we cannot refrain from reminiscing about one other
question, often heard in the field of dark energy: namely, ``Who
ordered this?". The pNGB quintessence models strive to be natural
in the sense of effective field theory, since the scales and
couplings governing their dynamics are radiatively stable. However
the pNGB dark energy with $f_a \la {\cal O}(1) M_{Pl}$, which were
a central subject of this study, do not appear all that natural or
simple from the point of view Occam's razor, a criterion often
employed to justify or belittle a model of dark energy. The
oft-quoted versions of Occam's razor are commonly extrapolated to
imply that dark energy may be the simplest of possibilities: a
cosmological constant. Yet this is a heuristic argument that does
not necessarily give correct answers. It is merely a loose guide
enabling one to pick a scientific hypothesis based on the fewest
currently unproven assumptions. To fully accept arguments based on
simplicity and Occam's razor, we need to go beyond na\"ive and
develop a {\it complete theory} of the dark sector, that goes
beyond the taxonomy of cosmic coincidences \cite{coincidence}.
Before we have it, we may be dismissing ideas too quickly. The
appearance of simplicity may be deceiving, as witnessed by a
dearth of complete theories of vacuum energy. Just putting a
number $\lambda$ in the equations is merely a fit. The pNGB
qui${\cal N}$tessence models are motivated by an attempt to
simultaneously address two theoretical issues: (1) protect the
ultra-low scales of quintessence from radiative corrections
involving the rest of the world\footnote{Except for the usual
vacuum energy cancellation, a problem which has to date defied all
attempts to solve it in common effective field theory in a natural
way \cite{weinberg}.}, and (2) derive ultra-weakly coupled
pNBG-like fields, simulating $f_{eff} > M_{Pl}$, from fundamental
theory. If it so happens that such models bring about a number of
observational consequences worth studying, all the better. Similar
opinions about Occam's razor have been expressed before on
numerous occasions (see \cite{occam} for a nice summary). We end
by quoting one such example:

\begin{flushleft}
{\it ~~~~~~~~~~~~~~~~``There are more things in heaven and earth, Horatio,}\\
{\it  ~~~~~~~~~~~~~~~~ Than are dreamt of in your philosophy."}\\
\hfill {\rm William Shakespeare}, {\it Hamlet}
~~~~~~~~~~~~~~~~~~~~~
\end{flushleft}

\noindent Whether or not it applies to our world remains to be
seen.

\vskip0.5cm

{\bf \noindent Acknowledgements}

\smallskip

We thank A.~Albrecht, H.-C.~Cheng, S.~Dimopoulos, M.~Kleban,
V.~Mukhanov, M.~Sloth, J.~Terning and J.~Wacker for useful
discussions. This work was supported in part by the DOE Grant
DE-FG03-91ER40674, in part by the NSF Grant PHY-0332258 and in
part by a Research Innovation Award from the Research Corporation.


\end{document}